\documentstyle[12pt,epsf]{article}

\begin{document}

\baselineskip=18.8pt plus 0.2pt minus 0.1pt

\makeatletter

\renewcommand{\thefootnote}{\fnsymbol{footnote}}
\newcommand{\beq}{\begin{equation}}
\newcommand{\eeq}{\end{equation}}
\newcommand{\bea}{\begin{eqnarray}}
\newcommand{\eea}{\end{eqnarray}}
\newcommand{\nn}{\nonumber}
\newcommand{\hs}[1]{\hspace{#1}}
\newcommand{\vs}[1]{\vspace{#1}}
\newcommand{\Half}{\frac{1}{2}}
\newcommand{\p}{\partial}
\newcommand{\ol}{\overline}
\newcommand{\wt}[1]{\widetilde{#1}}
\newcommand{\ap}{\alpha'}
\newcommand{\bra}[1]{\left\langle  #1 \right\vert }
\newcommand{\ket}[1]{\left\vert #1 \right\rangle }
\newcommand{\vev}[1]{\left\langle  #1 \right\rangle }

\newcommand{\ul}[1]{\underline{#1}}
\newcommand{\tr}{\mbox{tr}}
\newcommand{\ishibashi}[1]{\left\vert #1 \right\rangle\rangle }

\makeatother

\begin{titlepage}
\title{
\hfill\parbox{4cm}
{\normalsize KIAS-P01047\\{\tt hep-th/0110270}}\\
\vspace{1cm}
Comments on D-branes on general group manifolds
}
\author{Yoji Michishita
\thanks{
{\tt michishi@kias.re.kr}
}
\\[7pt]
{\it School of Physics, Korea Institute for Advanced Study}\\
{\it 207-43, Cheongryangri, Dongdaemun, Seoul, 130-012, Korea}
}

\date{\normalsize October, 2001}
\maketitle
\thispagestyle{empty}

\begin{abstract}
\normalsize
We investigate D-branes with maximal symmetry on general group manifolds
in terms of boundary states and effective actions.
We show that in large $k$ limit boundary states with 
an suitable Wilson line form 
boundary states of the other types of D-branes, extending the known 
fact in SU(2) case.
We also show that fluctuation mass spectrum around D-brane solutions of 
the effective action agrees with that of boundary CFT in large $k$ limit.
\end{abstract}



\end{titlepage}

\section{Introduction}

D-branes on flat space has been investigated extensively in the last 
several years, and various properties on D-branes on curved 
space has also been clarified gradually. 

For extending the analysis of D-branes to curved space, group manifolds
are good examples since CFTs on them i.e. WZW models are exactly solvable
and geometrical meaning is clearer than many other abstract CFTs. 
Recently D-branes on SU(2) have been studied extensively. 
One of the interesting phenomena is 
that D2-branes can be regarded as bound states of D0-branes\cite{bds,ars2}.

In this letter we investigate D-branes with maximal symmetry 
on general group manifolds in 
terms of boundary states and effective actions extending the work of 
SU(2) case. We will consider maximally symmetric D-branes given by Cardy
boundary states. In particular we consider formation of D-branes by 
gauge field condensation.

This letter is organized as follows.
In section 2 we briefly review D-brane on general group manifolds in 
terms of WZW models, and comment on the exponent of the characters 
in the case of D0-branes.
In section 3 we extend the boundary state analysis given in \cite{hns} 
to general group manifolds. In large $k$ limit 
($k$ is the level of the current algebra), 
we show that when we turn on a Wilson line 
on a bunch of D-branes it forms other types of D-branes.
Then we comment on supersymmetry.
In section 4 we consider effective actions given in \cite{ars2,fs}
, solutions of the equation 
of motion and fluctuation mass spectrum. We show that the spectrum 
agrees with that of CFT in large $k$ limit. 
We also comment on tachyon potentials.

\section{Maximally symmetric D-branes on general group manifolds}

When we consider D-branes in WZW models we must impose 
some boundary condition which relates the left moving current $J^a$
to the right moving current $\bar{J}^a$.
Here we consider the following condition which preserves the symmetry
of the current algebra (in terms of open strings).
\beq
J=r\bar{J}r^{-1},
\label{bc}
\eeq
where $r$ is an element of the group $G$, $J=J^aT^a$ and $T^a$ are 
generators of $G$. The action of $r$ on $\bar{J}$ is an inner 
automorphism. In addition to this one can consider outer automorphisms,
though we do not treat them in this paper. 

It is known that the D-brane with this boundary condition is wrapped 
on $C_h\cdot r$, where $C_h$ is the conjugacy class of $h$ \cite{as}.
Furthermore $h$ is `quantized' by quantum consistency.

The building blocks of boundary states are called Ishibashi states 
\cite{io},
which are constructed for each primary operator.
We denote the Ishibashi state for the boundary condition (\ref{bc}) 
constructed on the primary field 
belonging to the representation with the highest weight 
$\mu$ (We always denote representations of Lie algebra by 
their highest weights.) by $\ishibashi{\mu,r}$
Then $\ishibashi{\mu,r}$ is defined as follows.
\beq
\ishibashi{\mu,r}
\equiv\sum_n\ket{\mu,n}_L\otimes 
 e^{2\pi i\theta^a\bar{J_0}^a}U\ket{\mu,n}_R,
\eeq
where $\ket{\mu,n}$ are orthonormal bases of the module constructed on 
the representation $\mu$, and $\theta^a$ is defined by 
$r=e^{2\pi i\theta^a T^a}$. $U$ is an operator which satisfy 
$U^{-1}J_nU=-J^\dagger_{-n}$.

Then we can construct boundary states consistent with the Cardy 
condition by taking appropriate linear combination of 
$\left\vert \mu,r \right\rangle\rangle$. In the case of 
the diagonal modular invariant partition function they are given 
as follows\cite{c} .
\beq
\ket{\lambda,r}=\frac{S_\lambda^{\;\;\mu}}{\sqrt{S_\mu^{\;\; 0}}}
\ishibashi{\mu,r},
\eeq
where $S_\lambda^{\;\;\mu}$
is the modular transformation matrix of characters of current algebra
(See e.g. \cite{fms}):
\beq
\chi_\lambda(\zeta/\tau,-1/\tau)=S_\lambda^{\;\;\mu}
q^{\frac{k}{2}|\zeta|^2}\chi_\mu(\zeta,\tau).
\eeq
Characters $\chi_\lambda(\zeta,\tau)$ are defined as follows.
\beq
\chi_\lambda(\zeta,\tau)= \tr_\lambda\left[q^{L_0-\frac{c}{24}}
 \exp{(-2\pi i \zeta\cdot H)}\right],\quad q=e^{2\pi i\tau}.
\eeq
$H^i$ are generators of the Cartan subalgebra and $c$ is 
the central charge.

$S_\lambda^{\;\;\mu}$ has the following properties.
\beq
S^{-1} = S^\dagger,\quad ^tS=S,\quad 
(S_\lambda^{\;\;\mu})^* = S_{\lambda^*}^{\;\;\mu}
 = S_\lambda^{\;\;\mu^*},
\eeq
\beq
S_\lambda^{\;\;\mu} = 
 \wt{\chi}_\mu\left(\frac{-2\pi i}{k+g}(\lambda+\rho)\right)
 S_\lambda^{\;\; 0},
\eeq
where $\wt{\chi}_\mu$ is the character of ordinary Lie algebras.
$\lambda^*$ is the conjugate representation of $\lambda$.
$k$, $g$ and $\rho$ are level of the current algebra, 
the dual Coxeter number and the Weyl vector respectively.

We can read off the open string spectrum by computing the cylinder 
amplitude using these boundary states and modular transforming 
to the open string picture:
\bea
\bra{\lambda,r'}\wt{q}^{\Half(L_0+\bar{L}_0)-\frac{c}{24}}\ket{\mu,r}
& = & \frac{(S^\dagger)_\sigma^{\;\;\lambda} S_\mu^{\;\;\sigma}}
{S_\sigma^{\;\; 0}}\chi_\sigma(-\zeta,-1/\tau) \nn\\
 & = & \frac{S_{\lambda^*}^{\;\;\sigma} S_\mu^{\;\;\sigma} 
(S^\dagger)_\sigma^{\;\;\omega}}
{S_\sigma^{\;\; 0}}
q^{\frac{k}{2}|\zeta|^2}\chi_\omega(\tau\zeta,\tau) \nn\\
 & = & N_{\lambda^* \mu}^\omega 
q^{\frac{k}{2}|\zeta|^2}\chi_\omega(\tau\zeta,\tau),
\label{ospectrum}
\eea
where we used ${\left\langle\langle \lambda,r' \right\vert}
\wt{q}^{\Half(L_0+\bar{L}_0)-\frac{c}{24}}\ishibashi{\mu,r}
={\left\langle\langle \lambda,1 \right\vert}
\wt{q}^{L_0-\frac{c}{24}}e^{-2\pi i\theta'^aJ^a_0}e^{2\pi i\theta^aJ^a_0}
\ishibashi{\mu,1}
=\chi_\mu(-\zeta,-1/\tau)$, and the Verlinde formula \cite{v}.
$\zeta^a$ is defined by 
$e^{2\pi i\zeta^aT^a}=e^{-2\pi i\theta'^aT^a}e^{2\pi i\theta^aT^a}$. 
Here we assumed, without loss of generality,
$\zeta^a$ has nonzero value
only when the index $a$ is along the direction of Cartan subalgebra.
Notice that (\ref{ospectrum}) corresponds to open strings from 
$\ket{\mu,r}$ to $\ket{\lambda,r'}$. For open strings with 
the opposite orientation $\lambda, r'$ and $\mu, r$ are exchanged.

$\ket{0,r}$ corresponds to D0-brane.
For D0-branes conjugacy classes must be trivial. Therefore 
the position of $\ket{0,r}$ in $G$ is $r$. 

In general power of $q$ in the partition function is $\ap$ times
mass squared of modes of strings. Hence we can expect that the power 
of the factor $q^{\frac{k}{2}|\zeta|^2}$ in (\ref{ospectrum}) 
is equal to $\ap$ times
the square of the tension times
the length of geodesic connecting $r$ and $r'$. We will show that 
it is correct.

The metric on $G$ can be read off from the worldsheet 
action of the WZW model. We denote the parameter of the geodesic by 
$s$. $s=0$ and $s=1$ correspond to $r$ and $r'$ 
respectively. Then the length $L$ is 
\beq
L=\int_0^1 ds \sqrt{\left(-\frac{k\ap}{2}\right)
\tr[g^{-1}\p_s g g^{-1}\p_s g]}.
\eeq
The geodesic can be derived by minimizing $L$:
\beq
\p_s(\p_s gg^{-1})=0.
\eeq
The solution of this equation is 
\beq
g(s)=re^{-2\pi i\zeta s}
\label{geodesic}.
\eeq
Therefore,
\beq
L=2\pi\sqrt{\frac{k\ap}{2}|\zeta|^2}.
\eeq
Then $\ap(L/2\pi\ap)^2=\frac{k}{2}|\zeta|^2$, as expected.

\section{Wilson lines and boundary states}

In \cite{hns} it is shown that, in the SU(2) case, boundary states with
the following Wilson line operator satisfies the boundary condition
(\ref{bc}) again.
\beq
\mbox{Ptr}\exp\left(-\frac{i}{k}\int_0^{2\pi}d\sigma 
J^a(\sigma)M^a\right),\quad 
[M^a,M^b]=if^{abc}M^c.
\eeq
This fact is also true for general group manifolds.
Indeed, the argument given in \cite{hns} can be applied to 
the case of general group manifolds straightforwardly.
In addition to this, it is shown
that in the large $k$ limit we can get D2-brane 
boundary states by acting the above Wilson line operator 
on boundary state of D0-branes.

We will show this is also true in the case of general group by 
extending the calculation given in \cite{hns}.

First we determine how the Wilson line operator acts on each 
Ishibashi state. Since the Wilson line operator does not 
change boundary condition as mentioned above, and its action closes 
in each module, the result must be proportional to the Ishibashi state 
itself:
\beq
\mbox{Ptr}\exp\left(-\frac{i}{k}\int_0^{2\pi}d\sigma J^a(\sigma)M^a\right)
 \ishibashi{\lambda,r}
 = c(\mu,\lambda,k)\ishibashi{\lambda,r},
\eeq
where $M^a$ are representation matrices of $\mu$.
To extract $c(\mu,\lambda,k)$ we can concentrate only on 
the primary state part of the Ishibashi state.
In the large $k$ limit, we can remove the path ordering of the Wilson line.  
\beq
\tr\exp\left(-\frac{2\pi i}{k}J_0^aM^a\right)\ishibashi{\lambda,r}
 = \sum_{\mu'}\bra{\mu,\mu'}\exp\left(-\frac{2\pi i}{k}J_0^aM^a\right)
 \ket{\mu,\mu'}\otimes\ishibashi{\lambda,r},
\eeq
where $\mu'$ is a weight of $\mu$.
The operator $J_0^aM^a$ in the exponent can be diagonalized 
as follows.
\bea
J_0^aM^a & = & \Half[(J_0^a+M^a)^2-(J_0^a)^2-(M^a)^2] \nn\\
 & = & \Half[(\nu,\nu+2\rho)-(\lambda,\lambda+2\rho)-(\mu,\mu+2\rho)],
\eea
where we used the quadratic Casimir operator $Q_\lambda$ for a 
representation $\lambda$ is equal to $(\lambda,\lambda+2\rho)$.
$\nu$ is the representation coming from the tensor product of 
$\lambda$ and $\mu$.
Then $c(\mu,\lambda,k)$ can be read off as follows.
\beq
c(\mu,\lambda,k)=\sum_{\nu}{\cal N}_{\mu\lambda}^\nu
 \exp\left(-\frac{\pi i}{k}
 [(\nu,\nu+2\rho)-(\lambda,\lambda+2\rho)-(\mu,\mu+2\rho)]\right),
\eeq
where ${\cal N}_{\mu\lambda}^\nu$ is the tensor product coefficient, 
i.e. the multiplicity of the 
representation $\nu$ in the decomposition of the tensor product 
$\mu\otimes\lambda$. 
This is given by the following formula derived by the character method.
\beq
{\cal N}_{\mu\lambda}^\nu=
 \sum_{\stackrel{\mbox{$\scriptstyle 
 \mu'\in\{\mbox{\tiny weights of } \mu\}$}}
 {\mbox{$\scriptstyle w\in\mbox{\tiny Weyl group }, 
 w(\lambda+\mu'+\rho)-\rho=\nu\in\{\mbox{\tiny dominant weights}\}$}}}
 \epsilon(w) \mbox{mult}_\mu(\mu').
\label{tpcoeff}
\eeq
where $\epsilon(w)$ is the sign of $w$ and $\mbox{mult}_\mu(\mu')$
is the multiplicity of the weight $\mu'$ in the representation $\mu$.

We assume that only $|\lambda_i-\mu_i|\gg 1$ and $|\lambda_i|\gg 1$ 
contribute, where $\lambda_i$ and $\mu_i$
are the Dynkin labels of $\lambda$ and $\mu$.
Then, only the unit element of the Weyl group contribute to the sum
of (\ref{tpcoeff}), because $\lambda$, and $(\lambda+\mu'+\rho)$ by
the above assumption, are in the fundamental chamber, and 
$w(\lambda+\mu'+\rho)$, and $w(\lambda+\mu'+\rho)-\rho$ by the above 
assumption again, are outside the fundamental chamber for nontrivial 
$w$.
Therefore,
\beq
{\cal N}_{\mu\lambda}^\nu \sim \left\{
 \begin{array}{ll}
 \mbox{mult}_\mu(\mu') & \mbox{for}\quad \nu=\lambda+^\exists\mu', \\
 0 & \mbox{otherwise.} \\
 \end{array}\right.
\eeq
Now we can compute $c(\mu,\lambda,k)$:
\bea
c(\mu,\lambda,k) & \sim & \sum_{\mu'}\mbox{mult}_\mu(\mu')
 \exp\left(-\frac{\pi i}{k}
 [(\lambda+\mu',\lambda+\mu'+2\rho)-(\lambda,\lambda+2\rho)-(\mu,\mu+2\rho)]
 \right)
 \nn\\
 & = & \sum_{\mu'}\mbox{mult}_\mu(\mu')
 \exp\left(-\frac{\pi i}{k}
 [2(\lambda+\rho,\mu')-(\mu,\mu+2\rho)+(\mu',\mu')]\right)
 \nn\\
 & \sim & \sum_{\mu'}\mbox{mult}_\mu(\mu')
 \exp\left(-\frac{2\pi i}{k+g}(\lambda+\rho,\mu')\right)
 \nn\\
 & = & \wt{\chi}_\mu\left(-\frac{2\pi i}{k+g}(\lambda+\rho)\right)
 \nn\\
 & = &  \frac{S_\lambda^{\;\;\mu}}{S_\lambda^{\;\; 0}}.
\eea
Then we can determine the action of the Wilson line:
\bea
\mbox{Ptr}\exp\left(\frac{i}{k}\int_0^{2\pi}d\sigma J^a(\sigma)M^a\right)
\ket{\nu,r}
 & = & \mbox{Ptr}
 \exp\left(\frac{i}{k}\int_0^{2\pi}d\sigma J^a(\sigma)M^a\right)
\frac{S_\nu^{\;\;\lambda}}{\sqrt{S_\lambda^{\;\; 0}}}
\ishibashi{\lambda,r} \nn\\
 & = & 
\frac{S_\nu^{\;\;\lambda}}{\sqrt{S_\lambda^{\;\; 0}}}
\frac{S_\lambda^{\;\;\mu}}{S_\lambda^{\;\; 0}}
\ishibashi{\lambda,r} \nn\\
 & = & 
\frac{S_\nu^{\;\;\lambda}}{\sqrt{S_\lambda^{\;\; 0}}}
\frac{S_\mu^{\;\;\lambda}}{S_\lambda^{\;\; 0}}
(S^\dagger)_\lambda^{\;\;\xi}S_\xi^{\;\;\pi}
\ishibashi{\pi,r} \nn\\
 & = & 
N_{\nu\mu}^\xi
\frac{S_\xi^{\;\;\pi}}{\sqrt{S_\pi^{\;\; 0}}}
\ishibashi{\pi,r} \nn\\
 & = & 
N_{\nu\mu}^\xi\ket{\xi,r}.
\eea
This shows that by the action of the Wilson line D-branes $\ket{\nu,r}$ 
form other types of D-branes in large $k$ limit. 
The multiplicity of $\ket{\xi,r}$ is 
$N_{\nu\mu}^\xi$. In particular, $\ket{\nu,r}$ can be regarded as 
a bound state of dim $\nu$ D0-branes.

Finally we comment on supersymmetry.
In supersymmetric WZW models the current $J^a$ can be written as a sum of 
fermion part and remaining part independent of the fermion:
\beq
J^a=j^a-\frac{i}{2k}f^a_{\;\; bc}:\psi^b\psi^c:.
\eeq
If the index $a$ is in the Cartan subalgebra, then
\beq
J^i=j^i+\frac{1}{k}\sum_{\alpha>0}\alpha^i:\psi^\alpha\psi^{-\alpha}:,
\eeq
where $\alpha$s are positive roots. We can bosonize $\psi^\alpha$ as follows.
\beq
\psi^{\pm\alpha}=\frac{\sqrt{2k}}{|\alpha|}e^{\pm i\phi_\alpha},
\quad \phi_\alpha(z)\phi_\alpha(w)\sim-\ln(z-w),
\eeq
where we omit cocycle factors. Then, 
\beq
J^i=j^i+2i\sum_{\alpha>0}\frac{\alpha_i}{|\alpha|^2}\p\phi_\alpha.
\eeq
Spacetime supercharges $Q_{\epsilon_{\alpha_1}\epsilon_{\alpha_2}\dots}$ 
are written by spin operators $\exp(\frac{i}{2}
\sum_{\alpha>0}\epsilon_\alpha\phi_\alpha)$, where $\epsilon_\alpha=\pm 1$.
When the boundary state $\ket{\lambda,1}$($\times$ other part from 
the remaining part of the spacetime) preserves the following combination 
of left moving and right moving supersymmetries,
\beq
(Q_{\epsilon_{\alpha_1}\epsilon_{\alpha_2}\dots}
+\Lambda_{\epsilon_{\alpha_1}\epsilon_{\alpha_2}\dots}
 ^{\epsilon'_{\alpha_1}\epsilon'_{\alpha_2}\dots}
\bar{Q}_{\epsilon'_{\alpha_1}\epsilon'_{\alpha_2}\dots})\ket{\lambda,1}=0,
\eeq
then, noting $\ket{\lambda,r}=\exp(-i\theta^iJ_0^i)\ket{\lambda,1}$,
we can see $\ket{\lambda,r}$ preserves the following supersymmetry.
\beq
\left(Q_{\epsilon_{\alpha_1}\epsilon_{\alpha_2}\dots}
+\exp\left(i\sum_{\alpha>0}\epsilon_\alpha
 \frac{\theta^i\alpha_i}{|\alpha|}\right)
 \Lambda_{\epsilon_{\alpha_1}\epsilon_{\alpha_2}\dots}
 ^{\epsilon'_{\alpha_1}\epsilon'_{\alpha_2}\dots}
 \bar{Q}_{\epsilon'_{\alpha_1}\epsilon'_{\alpha_2}\dots}\right)
 \ket{\lambda,r}=0.
\eeq
As is pointed out in \cite{hns}, the sum of $\ket{\lambda,1}$ and 
$\ket{\lambda,r}$ does not preserve any supersymmetry in the case of SU(2).
This is because the factor 
$\exp(i\sum_{\alpha>0}\epsilon_\alpha\frac{\theta^i\alpha_i}{|\alpha|})$
becomes $\exp(i\epsilon\theta/2)$ in this case, and if this factor is equal 
to $1$, then $r=1$. 
However, if we consider group with the rank more than 2, we may put 
$\exp(i\sum_{\alpha>0}\epsilon_\alpha\frac{\theta^i\alpha_i}{|\alpha|})=1$
without putting $r=1$. For example, in the case of SU(3) 
$\theta=2\omega_1+\omega_2$ is such a choice, where $\omega_i$ are the 
fundamental weights. This fact shows that, in particular, two or more 
D0-branes put on different positions may be supersymmetric when we embed
WZW models with the rank more than 2 in superstring theory.

\section{Classical solutions of the effective action and 
mass spectrum of fluctuation around them}

We assume that the background is 
(flat time direction)$\times G\times$other 
part which we do not consider. 
The time direction is introduced only for measuring masses.
We consider only the state $J_{-1}^a\ket{0}$ in the case of 
bosonic string (and $\psi_{-1/2}^a\ket{0}$ in the case of superstring).
The effective action has been determined in \cite{ars2,fs} up to 
overall normalization:
\beq
S=\tr\left(\frac{2\ap}{k}(\p_tB_a)^2
  +\frac{1}{4k^2}[B_a,B_b]^2-\frac{i}{3k^2}f^{abc}B_a[B_b,B_c]\right).
\eeq
Here we do not distinguish upper and lower adjoint indices.
The trace is taken over Chan-Paton factor. $B_a$ are Hermitian matrices.
The power of $k$ for each term can be understood by noting that open 
string metric used for summation over adjoint indices contain $k$ 
\cite{ars2}.
$\ap$ in front of the time derivative comes from dimensional analysis.

The equation of motion for static configurations is
\beq
[B_b,[B_a,B_b]-if^{abc}B_c]=0.
\eeq
In \cite{ars2,fs} (See also \cite{cs}) 
it is pointed out that this equation has the solution
\beq
B_a=S_a,\quad [S_a,S_b]=if^{abc}S_c,
\eeq
and it represents the D-brane corresponding to the boundary state
$\ket{\lambda,1}$ if $S_a$ is the representation matrix of $\lambda$.
Direct sum of this type of solutions is also a solution and it represents
the configuration that various D-branes are present simultaneously.
In addition to these, 
\beq
B_a=\left(\begin{array}{cc} 
c_a & 0 \\
0 & S^a \end{array}\right),
\eeq
where $c_a$ is a constant, is also a solution and in \cite{hk} it is 
claimed that in the SU(2) case
this represents a D2-brane and a D0-brane. $c_a$ corresponds to the 
position of the D0-brane.

Here we consider the following solution.
\beq
B_a=S_a+c_a\cdot 1.
\eeq
We show that this solution corresponds to the boundary state 
$\ket{\lambda,r}$ by computing mass spectrum of fluctuations 
around it and comparing with that of CFT in large $k$ limit. 
$c_a$ corresponds to $r$. 
Similar calculations for SU(2) have been done in \cite{hk} and 
\cite{iktw,bt,jmwy} in different context.

Let us consider the direct sum of two such solutions:
\beq
B^0_a=\left(\begin{array}{cc} 
S^a+c_a\cdot 1 & 0 \\
0 & S'^a+c'_a\cdot 1 \end{array}\right),
\label{solution}
\eeq
where $S^a$ and $S'^a$ are generators of representations $\lambda$ and 
$\mu$ respectively. The low lying CFT spectrum of the 
configuration corresponding 
to this solution i.e. $\ket{\lambda,r}$ and $\ket{\mu,r'}$ can be read 
off from (\ref{ospectrum}). The results are shown in table 1 and 2.

\begin{table}[htbp]
\begin{tabular}{|c|c|c|}\hline
 & \multicolumn{2}{|c|}{$\ket{\mbox{primary}}$} \\ \cline{2-3}
 & representation & mass squared \\ \hline
open strings on & 
 $\sum_{\sigma}N_{\lambda\lambda^*}^\sigma\sigma$ & 
 $\frac{1}{\ap}\left(\frac{Q_\sigma}{2(k+g)}+\ap m^2\right)$ \\
$\ket{\lambda,r}$ & & \\ \hline
open strings on & 
 $\sum_{\sigma}N_{\mu\mu^*}^\sigma\sigma$ & 
 $\frac{1}{\ap}\left(\frac{Q_\sigma}{2(k+g)}+\ap m^2\right)$ \\
$\ket{\mu,r'}$ & & \\ \hline
open strings between & 
 $\sum_{\sigma}N_{\lambda\mu^*}^\sigma\sigma$ & 
 $\frac{1}{\ap}\left(\frac{Q_\sigma}{2(k+g)}-\zeta_i\sigma'_i
 +\frac{k}{2}|\zeta|^2+\ap m^2\right)$ \\
$\ket{\lambda,r}$ and $\ket{\mu,r'}$ & 
 $+\sum_{\sigma}N_{\mu\lambda^*}^\sigma\sigma$ & 
 $\sigma'$ : weights of $\sigma$
  \\ \hline
\end{tabular}
\caption{representations and mass spectrum of the states 
$\ket{\mbox{primary}}$}
\end{table}

\begin{table}[htbp]
\begin{tabular}{|c|c|c|}\hline
 & \multicolumn{2}{|c|}{$J_{-1}^a\ket{\mbox{primary}}$ or 
 $\psi_{-1/2}^a\ket{\mbox{primary}}$} \\ \cline{2-3}
 & representation & mass squared \\ \hline
open strings on & 
 $\sum_{\sigma}N_{\lambda\lambda^*}^\sigma\sigma\otimes$(adjoint) & 
 $\frac{1}{\ap}\left(\frac{Q_\sigma}{2(k+g)}\right)$ \\
$\ket{\lambda,r}$ & & \\ \hline
open strings on & 
 $\sum_{\sigma}N_{\mu\mu^*}^\sigma\sigma\otimes$(adjoint) & 
 $\frac{1}{\ap}\left(\frac{Q_\sigma}{2(k+g)}\right)$ \\
$\ket{\mu,r'}$ & & \\ \hline
open strings between &
 $\sum_{\sigma,\nu}N_{\lambda\mu^*}^\sigma N_{\sigma,
 \mbox{\tiny adj.}}^\nu\nu$ & 
 $\frac{1}{\ap}\left(\frac{Q_\sigma}{2(k+g)}-\zeta_i\nu'_i
 +\frac{k}{2}|\zeta|^2\right)$ \\
$\ket{\lambda,r}$ and $\ket{\mu,r'}$ & 
 $+\sum_{\sigma,\nu}N_{\mu\lambda^*}^\sigma N_{\sigma,
 \mbox{\tiny adj.}}^\nu\nu$ & 
 $\nu'$ : weights of $\nu$
  \\ \hline
\end{tabular}
\caption{representations and mass spectrum of the states 
$J_{-1}^a\ket{\mbox{primary}}$ (bosonic string) or 
 $\psi_{-1/2}^a\ket{\mbox{primary}}$ (superstring)}
\end{table}

Quadratic Casimir operators in the table 1 and 2
come from dimensions of primary operators.
In superstring case $k$ is shifted to $k-g$.
$m^2$ in the table 1 is the contribution of zero point energy.
It is given by $m^2=-\frac{1}{\ap}$ 
in the case of bosonic D-branes and 
$m^2=-\frac{1}{2\ap}$ in the case of non-BPS 
D-branes in superstring theory.

Here we note the open strings from $\ket{\mu,r'}$ to $\ket{\lambda,r}$
give complex conjugate representations of 
the open strings from $\ket{\lambda,r}$ to $\ket{\mu,r'}$ and, however,
they give the identical mass spectrum since the sign of 
both weights and $\zeta$ flip. 

If we plug $B_a=B^0_a+\delta B_a$ into the action and
 calculate the part quadratic
in the fluctuation $\delta B_a$, then we obtain
\bea
\delta^2 S & = & \frac{\ap}{k}\tr\left((\p_t\delta B_a)^2\right) \nn\\
 & & +\frac{1}{k^2}\tr\left(\Half[B_a^0,\delta B_b]^2
 -\Half [B_a^0,\delta B_a]^2
 +[B_a^0,B^0_b][\delta B_a,\delta B_b]
 -if^{abc}B_a^0[\delta B^0_b,\delta B^0_c]\right).
\label{quadfluc1}
\eea
We take the following as gauge fixing terms \cite{iktw}.
\beq
\frac{1}{k^2}\tr\left(\Half[B_a^0,\delta B^a]^2+[B_a^0,b][B_a^0,c]\right).
\label{gaugefix}
\eeq
$b$ and $c$ are antighost and ghost respectively. Since 
we are interested in only physical spectrum, we drop the 
ghost term henceforth. The first term of (\ref{gaugefix}) cancels
the second term in the second trace of (\ref{quadfluc1}).
If we put $\delta B_a$ as follows,
\beq
\delta B_a=\left(\begin{array}{cc} 
D_a & E_a \\
E^\dagger_a & F_a
\end{array}\right),
\eeq
and use the explicit form of $B^0_a$, we obtain
\bea
\delta^2 S & = & \frac{\ap}{k}
 \tr\left((\p_tD_a)^2+\frac{1}{2\ap k}[S_a,D_b]^2\right) \nn\\
 & & +\frac{\ap}{k}
 \tr\left((\p_tD_a)^2+\frac{1}{2\ap k}[S'_a,F_b]^2\right) \nn\\
 & & \frac{2\ap}{k}\tr\left(\p_tE_a\p_tE^\dagger_a\right)
 +\frac{1}{k^2}
 \tr\Bigl((S'_a E^\dagger_b-E^\dagger_b S_a)(S_a E_b-E_b S'_a) 
 \nn\\
 & & -2(c_a-c'_a)E^\dagger_b(S_a E_b-E_b S'_a)
 +2if^{abc}(c_a-c'_a)E^\dagger_b E_c
 \nn\\
 & &  -(c_a-c'_a)^2 E^\dagger_b E_b\Bigr).
\label{quadfluc2}
\eea

$D_a$, $F_a$ and $E_a$ correspond to strings on 
$\ket{\lambda,r}$, on $\ket{\mu,r'}$ and betweeen $\ket{\lambda,r}$ and 
$\ket{\mu,r'}$, respectively.
Let us consider the part containing $D_a$. 
We define matrices $N_a$ as follows.
\bea
 [S_a,D_b]_i^j & = & (S_a)_i^k(D_b)_k^j-(D_b)_i^k(S_a)_k^j \nn\\
 & = & [(S_a)_i^k\delta_l^j-\delta_i^k(S_a^*)_l^j](D_b)_k^l \nn\\
 & \equiv & (N_aD_b)_i^j.
\eea
Indices $i,j,k,l$ run from $1$ to dim$\lambda$. We can 
expand vectors with these indices by the basis of 
(dim $\lambda$)-dimensional representation $\lambda$ of $G$.
Then we can think of $N_a$ as generators of the tensor product 
representation $\lambda\otimes\lambda^*$.
The mass term of $D_a$ can be rewritten as follows.
\bea
\tr\left(\frac{1}{2\ap k}[S_a,D_b]^2\right) & = & 
 \tr\left(-\frac{1}{2\ap k}D_b[S_a,[S_a,D_b]]\right) \nn\\
 & = & -\frac{1}{2\ap k}(D_b)_j^i((N_a)^2D_b)_i^j.
\eea
$(N_a)^2$ is the quadratic Casimir operator. We can diagonalize
this operator by decomposing $\lambda\otimes\lambda^*$ into 
$\sum_\sigma{\cal N}_{\lambda\lambda^*}^\sigma\sigma$.
Therefore $\mbox{dim}G\cdot{\cal N}_{\lambda\lambda^*}^\sigma
\cdot\mbox{dim}\sigma$ modes have mass squared $\frac{Q_\sigma}{2\ap k}$. 
(The factor $\mbox{dim}G$ comes from the index $b$ of $D_b$.)
This spectrum agrees with that given by CFT (table 2) 
in large $k$ limit,
since $k+g\sim k$ and ${\cal N}_{\lambda\lambda^*}^\sigma\sim 
N_{\lambda\lambda^*}^\sigma$.
Similarly the mass spectrum of $F_b$ agrees with the CFT.

Next we compute the mass spectrum of $(E_a)_i^{i'}$.
We can expand vectors with the index $i$ and $i'$ by the bases of 
$\lambda$ and $\mu^*$ respectively.
If we define $L_a$ by $(S_a E_b-E_b S'_a)_i^{i'}\equiv(L_aE_b)_i^{i'}$,
$L_a$ can be regarded as the generators of $\lambda\otimes\mu^*$.
Then,
\bea
\delta^2 S & = & \frac{2\ap}{k}\Biggl(
 \p_t(E^\dagger_a)_{i'}^i\p_t(E_a)_i^{i'}
 -\frac{1}{2\ap k}\Bigl[(E^\dagger_b)_{i'}^i((L_a)^2E_b)_i^{i'} \nn\\
 & & +2(c_a-c'_a)(E^\dagger_b)_{i'}^i((L_a)E_b)_i^{i'}
 -2if^{abc}(c_a-c'_a)(E^\dagger_b)_{i'}^i (E_c)_i^{i'}
 \nn\\
 & &  +(c_a-c'_a)^2 (E^\dagger_b)_{i'}^i (E_b)_i^{i'}\Bigr]\Biggr).
\label{equad}
\eea
Noting that $if^{abc}\equiv(T_{\mbox{\tiny adj.}}^b)_{ac}$ is generators
of adjoint representation, the second line of (\ref{equad})
can be written by $J^a\equiv L_a+T_{\mbox{\tiny adj.}}^a$:
\beq
 2(c_a-c'_a)(E^\dagger_b)_{i'}^i((L_a)E_b)_i^{i'}
 -2if^{abc}(c_a-c'_a)(E^\dagger_b)_{i'}^i (E_c)_i^{i'}
 =2(c_a-c'_a)(E^\dagger_b)_{i'}^i(J_a)_{bij'}^{ci'j}(E_c)_j^{j'}.
\eeq
Without loss of generality, we can assume that $c_a-c'_a$ has nonzero value
only when the index $a$ is along the direction of Cartan subalgebra.

Now we can read off the mass spectrum. The second term of (\ref{equad})
can be diagonalized in the same way as in the case of $D_b$:
$\mbox{dim}G\cdot{\cal N}_{\lambda\mu^*}^\sigma
\cdot\mbox{dim}\sigma$ modes have the eigenvalue $\frac{Q_\sigma}{2\ap k}$.
The third and fourth term can be diagonalized by decomposing 
$\lambda\otimes\mu^*\otimes\mbox{(adjoint)}$ into irreducible 
representations. $J_a$ gives their weights because of the above
assumption. Therefore $2\cdot{\cal N}^\sigma_{\lambda\mu^*}$ 
modes (The factor 2 is present because each component of $E_b$ is complex.) 
have the eigenvalue $2(c_i-c'_i)\nu'_i$, where $\nu$ is a weight
of $\sigma\otimes\mbox{(adjoint)}$.
The fifth term is already diagonalized.
Then the total eigenvalues $\frac{1}{\ap}[\frac{Q_\sigma}{2k}
+\frac{1}{k}(c_i-c'_i)\nu'_i+\frac{1}{2k}(c_i-c'_i)^2]$
is exactly the same as that obtained by CFT (table 2) 
in the large $k$ limit, 
if we identify $(c_i-c'_i)$ with $-k\zeta_i$.

Next let us include the tachyon field in the effective action.
The quadratic part of the action of the tachyon field is as follows.
\beq
S_T=\tr\left(\ap(\p_tT)^2+\frac{1}{2k}[B_a,T]^2-
\ap m^2T^2\right).
\eeq
$m^2T^2$ is the mass term of $T$. The term $[B_a,T]$ comes from the 
kinetic term ``$(D_aT)^2$''.
We consider the fluctuation around the solution (\ref{solution}) 
and $T=0$. We put the fluctuation of $T$ as follows.
\beq
\delta T=\left(\begin{array}{cc} 
X & Y \\
Y^\dagger & Z \end{array}\right).
\eeq
Then the quadratic part in $\delta T$ is
\bea
\delta^2 S_T & = & \ap\tr\left((\p_tX)^2
 +\frac{1}{2\ap k}[S_a,X]^2-m^2X^2\right) 
 \nn\\
 & & \ap\tr\left((\p_tZ)^2
 +\frac{1}{2\ap k}[S'_a,Z]^2-m^2Z^2\right)
 \nn\\
 & & +2\ap\tr(\p_tY^\dagger\p_tY)
 +\frac{1}{k}\tr\Bigl((S'_aY^\dagger-Y^\dagger S_a)(S_aY-YS'_a)  \nn\\
 & &  -2(c_a-c'_a)Y^\dagger(S_aY-YS'_a)
 -(c_a-c'_a)^2Y^\dagger Y\Bigr) -\tr(2\ap m^2Y^\dagger Y) \nn\\
 & = & \ap\left((\p_tX)_j^i(\p_tX)_i^j
 -\frac{1}{2\ap k}X_j^i((N_a)^2 X)_i^j-m^2X_j^iX_i^j\right) \nn\\
 & &  +\ap\left((\p_tZ)_{j'}^{i'}(\p_tZ)_{i'}^{j'}
 -\frac{1}{2\ap k}Z_{j'}^{i'}((N'_a)^2Z)_{i'}^{j'}
 -m^2Z_{j'}^{i'}Z_{i'}^{j'}\right)
 \nn\\
 & & +2\ap\Biggl((\p_tY^\dagger)_{i'}^i(\p_tY)_i^{i'}
 -\frac{1}{2\ap k}(Y^\dagger)_{i'}^i((L_a)^2Y)_i^{i'} \nn\\
 & &  -\frac{1}{\ap k}(c_a-c'_a)(Y^\dagger)_{i'}^i(L_aY)_i^{i'}
 -\frac{1}{2\ap k}(c_a-c'_a)^2(Y^\dagger)_{i'}^i Y_i^{i'} 
 -m^2(Y^\dagger)_{i'}^i Y_i^{i'}\Biggr).
\eea
We can read off the mass spectrum in the same way as in the previous case:
\bea
X & : & N_{\lambda\lambda^*}^\sigma \mbox{modes with mass squared } 
 \frac{1}{\ap}\left(\frac{Q_\sigma}{2k} +\ap m^2\right) \nn\\
Z & : & N_{\mu\mu^*}^\sigma \mbox{modes with mass squared } 
 \frac{1}{\ap}\left(\frac{Q_\sigma}{2k} +\ap m^2\right) \nn\\
Y & : & 2 N_{\lambda\mu^*}^\sigma \mbox{modes with mass squared } 
 \frac{1}{\ap}\left(\frac{Q_\sigma}{2k}+\frac{1}{k}(c_i-c'_i)\sigma'_i
 +\frac{1}{2k}(c_i-c'_i)^2 +\ap m^2\right)
\eea
These spectra agree with those given by CFT in large $k$ limit
(table 1).

Finally we comment the form of the tachyon potential.

If we consider only the state $\ket{0}$, the calculation of the potential
by BSFT \cite{bsft} can be done trivially.
\beq
V(T)=\left\{\begin{array}{ll}
 \tr(e^T(1+T)) & \mbox{for bosonic D-branes,} \\
 \tr(e^{-\frac{1}{4}T^{\dagger}T})
 + \tr(e^{-\frac{1}{4}TT^{\dagger}}) & \mbox{for brane-antibrane pairs,} \\
 \tr(e^{-\frac{1}{4}T^2}) & \mbox{for non-BPS D-branes.} \\
\end{array}\right.
\eeq
Note that the tachyon fields in BSFT are different from $T$ 
in the previous discussion. They are related by some nonlocal field 
redefinition.
In the case of large $k$ limit of SU(2) including nontrivial
primary operators, see \cite{io2}. See also \cite{cs}.

The above forms of tachyon potentials are correct not only on 
group manifolds, but also on general
backgrounds. Furthermore they are exact regardless of the value of $k$.
We can discuss vanishing of D-branes by tachyon condensation 
in the same way as on flat background.

\vs{.5cm}
\noindent
{\large\bf Acknowledgments}\\[.2cm]
I would like to thank M.\ Nozaki and S.-J.\ Sin for helpful 
correspondence and conversation.

\newcommand{\J}[4]{{\sl #1} {\bf #2} (#3) #4}
\newcommand{\andJ}[3]{{\bf #1} (#2) #3}
\newcommand{\AP}{Ann.\ Phys.\ (N.Y.)}
\newcommand{\MPL}{Mod.\ Phys.\ Lett.}
\newcommand{\NP}{Nucl.\ Phys.}
\newcommand{\PL}{Phys.\ Lett.}
\newcommand{\PR}{Phys.\ Rev.}
\newcommand{\PRL}{Phys.\ Rev.\ Lett.}
\newcommand{\PTP}{Prog.\ Theor.\ Phys.}
\newcommand{\hepth}[1]{{\tt hep-th/#1}}



\begin{thebibliography}{99}

\bibitem{bds}
 C.\ Bachas, M.\ Douglas and C.\ Schweigert, 
 ``Flux stabilization of D-branes'',
 \hepth{0003037}, \J{JHEP}{0005}{2000}{048}

\bibitem{ars2}
 A.\ Yu.\ Alekseev, A.\ Recknagel and V.\ Schomerus, 
 ``Brane Dynamics in Background Fluxes and Non-commutative Geometry'',
 \hepth{0003187}, \J{JHEP}{0005}{2000}{010}

\bibitem{hns}
 Y.\ Hikida, M.\ Nozaki and Y.\ Sugawara,
 ``Formation of spherical D2-brane from multiple D0-branes'',
 \hepth{0101211}

\bibitem{fs}
 S.\ Fredenhagen and V.\ Schomerus,
 ``Branes on group manifolds, gluon condensates, and twisted K-theory'',
 \hepth{0012164},\J{JHEP}{0104}{2001}{007}

\bibitem{as}
 A.\ Yu.\ Alekseev and V.\ Schomerus,
 ``D-brane in the WZW model'',
 \hepth{9812193}, \J{\PR}{D60}{1999}{061901}

\bibitem{io}
 N.\ Ishibashi, 
 ``The boundary and crosscap states in conformal field theories'',
 \J{\MPL}{A4}{1989}{251} ;
 N.\ Ishibashi and T.\ Onogi, ``Conformal field theories on
 surfaces with boundaries and crosscaps'',
 \J{\MPL}{A4}{1989}{161}

\bibitem{c}
 J.\ L.\ Cardy,
 ``Boundary conditions, fusion rules and Verlinde formula'',
 \J{\NP}{B324}{1989}{581}

\bibitem{fms}
 P.\ di Francesco, P.\ Mathieu and D.\ S\'{e}n\'{e}chal, 
 ``Conformal Field Theory'',
 Springer

\bibitem{v}
 E.\ Verlinde,
 ``Fusion rules and modular transformations in 2D conformal field theory'',
 \J{\NP}{B300}{1988}{360}

\bibitem{cs}
 L.\ Cornalba and R.\ Schiappa,
 ``Nonassociative star product deformations for D-brane 
 worldvolumes in curved backgrounds'',
 \hepth{0101219}

\bibitem{hk}
 K.\ Hashimoto and K.\ Krasnov,
 ``D-brane solutions in non-commutative gauge theory on fuzzy sphere'',
 \hepth{0101145}, \J{\PR}{D64}{2001}{046007}

\bibitem{iktw}
 S.\ Iso, Y.\ Kimura, K.\ Tanaka and K.\ Wakatsuki,
 ``Noncommutative gauge theory on fuzzy shpere from matrix model'',
 \hepth{0101102}, \J{NP}{B604}{2001}{121}

\bibitem{bt}
 S.\ Bal and H.\ Takata,
 ``Interaction between two fuzzy spheres'',
 \hepth{0108002}

\bibitem{jmwy}
 D.\ P.\ Jatkar, G.\ Mandal, S.\ R.\ Wadia and K.\ P.\ Yogendran,
 ``Matrix dynamics of fuzzy spheres'',
 \hepth{0110172}

\bibitem{bsft}
A.\ A.\ Gerasimov and S.\ L.\ Shatashvili,
``On exact tachyon potential in open string field theory'',
\J{JHEP}{0010}{2000}{034}, \hepth{0009103} ;
D.\ Kutasov, M.\ Mari\~{n}o and G.\ Moore,
``Some exact results on tachyon condensation in string field theory'',
\J{JHEP}{0010}{2000}{045}, \hepth{0009148} ;
D.\ Kutasov, M.\ Mari\~{n}o and G.\ Moore,
``Remarks on tachyon condensation in superstring field theory'',
 \hepth{0010108}

\bibitem{io2}
 H.\ Ita and Y.\ Oz,
 ``String field theory and the fuzzy sphere'',
 \hepth{0106187}

\end{thebibliography}
\end{document}